\documentclass[preprint]{aastex}
\usepackage{psfig} 

\def\kms{~km~s$^{-1}$}
\def\kmsb{km~s$^{-1}$}
\def\hal{H$\alpha$}
\def\be{\begin{equation}}
\def\ee{\end{equation}}

\def\h{~$h_{100}^{-1}$}
\def\ha{~$h_{70}^{-1}$}

\def\HI{\hbox{H \sc i}}

\def\NII{[\ion{N}{2}]}

\def\SII{[\ion{S}{2}]}

\def\Ropt{$R_{\rm opt}$}
\def\Rcut{$R_{\rm cut}$}

\def\p{$\pm$}

\begin {document}
\slugcomment{\scriptsize \today \hskip 0.2in Version 2.5}

\title{Signatures of Galaxy-Cluster Interactions: Tully-Fisher Observations at
$z\sim0.1$}
 
\author{Daniel A. Dale}
\affil{Department of Physics and Astronomy, University of Wyoming, Laramie,
WY 82071}
\and
\author{Juan M. Uson}
\affil{National Radio Astronomy Observatory, 520 Edgemont Road,
Charlottesville, VA 22903}

\begin {abstract}
We have obtained new optical imaging and spectroscopic observations of
78~galaxies in the fields of the rich clusters  Abell~1413 ($z = 0.14$),
Abell~2218 ($z = 0.18$) and Abell~2670 ($z = 0.08$).  We have detected line
emission from 25 cluster galaxies plus an additional six galaxies
in the foreground and background, a much lower success rate than what was
found ($65$\%) for a sample of 52 lower-richness Abell clusters in the
range $0.02 \lesssim z \lesssim 0.08$.    We have combined these data with our
previous observations of Abell~2029 and Abell~2295 (both at $z = 0.08$), which
yields a sample of 156 galaxies.  We evaluate several parameters as a function
of cluster environment:  Tully-Fisher residuals, \hal\ equivalent width, and
rotation curve asymmetry, shape and extent.  Although \hal\ is more easily
detectable in galaxies that are located further from the cluster cores, we fail
to detect a correlation between \hal\ extent and galaxy location in those where
it is detected, again in contrast with what is found in the clusters of lesser
richness. We fail to detect any statistically significant trends for the other
parameters in this study.  The zero-point in the $z\sim0.1$ Tully-Fisher
relation is marginally fainter (by 1.5$\sigma$) than that found in nearby
clusters, but the scatter is essentially unchanged.  

\end {abstract}
 

\section {Introduction}

Galaxies in dense clusters evolve mainly due to tidal interactions, mergers and
ram pressure stripping with accretion contributing to the buildup of the cD
galaxies that often reside at the bottom of the cluster potential
(Dressler~1984).  These processes are expected to be most effective in the
cluster cores where higher densities will lead to more efficient ram pressure
stripping and tidal effects should be stronger.  Indeed, observations of
\hal\ in Abell~2029 failed to detect emission-line galaxies projected within 600\ha~kpc
of the center of the cluster, with the exception of a background galaxy, whereas
two-thirds of the galaxies located outside this radius were detected (Dale \&
Uson~2000).

It is well established that galaxies in clusters tend to be deficient in their
neutral Hydrogen (\HI) emission (e.g.~Haynes, Giovanelli \& Chincarini~1984).
More recently, Solanes et al.~(2001) analyzed a sample of 1900~galaxies in the
fields of eighteen nearby clusters ($z\lesssim0.04$) and found that \HI\
deficient galaxies are more likely located near the cluster cores.  This trend
holds true in Abell~2670, which is one of the most distant clusters to be
imaged in neutral Hydrogen ($z\sim0.08$, van Gorkom~1996).  A truly extreme
case is that of Abell~2029 which is at the same redshift as Abell~2670 but in
which a deeper search for \HI\ by one of us (JMU) resulted in an order of
magnitude lower detection rate.  \HI\ deficiency in cluster galaxies is usually
attributed to ram pressure stripping: as galaxies approach the central regions
of clusters at a speed $v$, they experience a strong
interstellar-gas--intracluster-gas interaction which is proportional to $v^2$.
This interaction is likely to induce a burst of star formation and subsequently 
strip away most of the the remaining
galactic interstellar gas (Fujita~1998; Fujita \& Nagashima~1999; Balogh et
al.~1999; Quilis, Moore \& Bower~2000).  

New results from the Sloan Digital Sky Survey show cluster effects that cannot
be reconciled by the morphology-density relation alone.  Using a volume-limited
sample of 8598 galaxies with $0.05 < z < 0.095$, Gomez et al. (2003) find that
the galaxy star formation rate decreases with increasing galaxy density.  In a
study of the rotation curves of 510~galaxies in 53~clusters spread throughout
$0.02\lesssim z \lesssim 0.08$, Dale et al.~(2001) found support for this
galaxy-cluster interaction scenario in the normalized radial extent of \hal\
emission in cluster spiral galaxies: it increases 4$\pm$2\% per Mpc of
(projected) separation of the galaxy from the cluster-core.  Analysis of the
kinematical asymmetries of the galaxies also shows a trend with cluster-centric
distance---rotation curve asymmetry is greater by a factor of two for inner
cluster early-type spiral galaxies.  These two results lead these authors 
to claim that {\it ``such trends are consistent with spiral disk perturbations
or even the stripping of the diffuse, outermost gaseous regions within the
disks as galaxies pass through the dense cluster cores''} (Dale et al.~2001).  

Are such observables different in $z \sim 0.1$ clusters?  Are they different
for richer, more X-ray
luminous clusters, thus solidifying the interpretation that the intracluster
environment can significantly impact galaxy evolution in clusters?  In an
effort to explore the Tully-Fisher relation at $z\sim0.1$, we have undertaken
a new imaging and spectroscopic campaign of four rich Abell clusters:
Abell~1413, Abell~2029, Abell~2218, and Abell~2670.  Results from the first
cluster surveyed, Abell~2029, were presented in Dale \& Uson~(2000).  Here, we
present a full analysis of the data for all four clusters.  We also include
results from Abell~2295, the only $z\sim0.08$ cluster observed in the Dale et
al.~(1999) survey.

\section {The Sample}

Our sample consists of four Abell clusters that are among the densest and
richest in the Abell catalog, thus providing good laboratories in which to
study the effects of the intracluster medium  and a fifth one (Abell~2295) of
similar redshift but much lower richness (Table~\ref{tab:cluster_sample}).  We
first discuss the three clusters for which new data have been obtained, and
then the two clusters for which our data have already been published.

{\bf Abell~1413}: Abell 1413 is a spiral rich cluster which harbors a dense
core of galaxies and an extremely large cD galaxy at the bottom of the cluster
potential well.  Indeed, recent $R$ band photometry shows that the cD halo
extends over 1~Mpc from the cluster center (Uson and Boughn, in preparation).
The average redshift is c$z_\odot = 42453 \pm 570$\kms\ based on 9 measurements,
and the velocity dispersion is $\sigma_{\rm 1d,rest} = 1490 $\kms\ (after
correction for the cosmological broadening), which is highly uncertain given
the small number of redshifts available for this cluster.  Abell (1958)
estimated its distance to be larger than it has turned out to be, because the
stripping of the outer envelopes of its galaxies due to their presumed
evolution in the dense environment of this cluster has resulted in a fainter
magnitude for the tenth-ranked galaxy, used by Abell to assign distances and
counting radii.  Given its redshift, Dressler~(1978) repeated the ``Abell
count'' and upgraded its richness to 4.5 (Abell estimated $R=3$).  A similar
increase in the richness of Abell~2029 and Abell~2218 is noted below, as is a
decrease in the derived richness of Abell~2670.  Bad weather severely limited
our data collection on Abell~1413.  Nevertheless, it seemed to lack galaxies
with useful \hal\ emission---observations of several galaxies in this cluster
yielded only two weak emission line detections.

{\bf Abell~2218}: The most distant cluster in the sample is Abell~2218 at
$z\sim0.18$, a dense cluster (Dressler's $R = 4.3$ vs. Abell's $R = 4$) also quite
rich in spiral galaxy candidates.  Extensive photometric and spectroscopic
studies have been performed on the gravitational lensing arclets near the core
of this cluster, contributing to excellent mass models of the region (Kneib et
al.~1996).  Deep $VRI$ mosaic images of this cluster detect diffuse
intracluster light at the $0.5\times 10^{9}~M_{\sun}~h_{70}^2~{\rm kpc}^{-2}$
level which traces the dark matter in the cluster (Boughn et al.~2000).  We
used the redshifts available in NED to compute a velocity dispersion
$\sigma_{\rm 1d,rest} \sim 2200$\kms\ which seemed too high.  Indeed, we have found that a number of the
redshifts that appear in NED are not the standard c$z$, but are instead the
relative velocities that would result in the measured spectra in a laboratory
experiment; i.e., deduced using the special-relativistic Doppler expression,
although this was not stated in the original reference (Ziegler et al.~2001).
The expression for the special-relativistic Doppler effect should not be used
over cosmological distances, because even if the Universe was ``flat'' it would
only be so in one (comoving) time slice.  Space-time is curved and relative
velocities are ill-defined over cosmological distances.  Observed quantities
should be reported, which in this case should be c$z$ (preferably
heliocentric).  Thus, we have corrected the Ziegler et al.~(2001)
``velocities'' to c$z$ and, after removing some duplications, we have derived
an average redshift of c$z_{\odot} = 52514 \pm 170$\kms\ and a velocity 
dispersion of $\sigma_{\rm 1d,rest} = 1437 \pm 100 $\kms\ with 93~redshifts
within one Abell radius.  This dispersion is reasonably consistent with the
value of 1202\kms\ derived by Kneib et al.~(1995) for the central region from
the lens model of the arcs in Abell~2218.

{\bf Abell~2670}: Abell~2670 is dominated by one of the largest cD galaxies
known.  Its richness was lowered by Dressler~(1978) from $R = 3$ (Abell~1958) to
$R = 1.4$.  A large number of redshifts have been measured for this cluster,
providing an average of c$ z_\odot = 22841 \pm 60$ and a velocity dispersion of
$\sigma_{\rm 1d,rest} = 900\pm40$\kms\ based on 235 galaxies located within a
radius of 3.6\ha~Mpc of the cluster center.  It has been noted that the cluster
shows significant substructure and might still be merging (Bird~1994).  A deep
VLA study of Abell~2670 detected $\sim$30 \HI\ sources, although none is
located near the cluster core (van Gorkom~1996).  Emission line spectroscopy
in the relatively \HI-rich Abell~2670 provides an important contrast to similar
work in \HI-poor Abell~2029.

{\bf Abell~2029}: Abell~2029 is one of the densest and richest clusters in
the Abell catalog.  The cluster is a textbook example of a compact, relaxed,
cD~cluster with high intracluster X-ray luminosity.  The cD~galaxy is one
of the largest galaxies known, with low surface brightness emission detected
out to a radius of 0.9\ha~Mpc (Uson, Boughn \& Kuhn~1991).  The average
redshift is c$z_\odot = 23220 \pm 170$\kms\ based on 92 measurements of galaxies
within 2.1\ha~Mpc and the velocity dispersion is
$\sigma_{\rm 1d,rest} = 1471 \pm 100$\kms\ (we have corrected the value reported 
in Dale \& Uson~2000 which was inadvertently lowered by an extra factor of
$1+z$).

{\bf Abell~2295}: This field is spiral rich, comprised of two superimposed
clusters well separated in redshift.  The farther cluster ($z\sim0.08$) is
dubbed ``Abell 2295,'' while the nearer cluster ($z\sim0.06$) is referred to
as ``Abell 2295b'' in Dale et al.~(1998).  Although of low richness (Abell's
$R=0$) it is the only cluster at a similar redshift to those in our sample
for which \hal\ rotation curves are available and it is included in this study
for comparison with our sample.  The average redshift is
c$z_\odot = 24623 \pm 200$\kms\ based on 9 measurements of galaxies within
2.1\ha~Mpc and the velocity dispersion is $\sigma_{\rm 1d,rest} = 560\pm160$\kms.

\section {Observations and Data Reduction}
\label{sec:data}

The observing techniques and data reduction for Abell~2295 and Abell~2029 are
described in Dale et al.~(1998) and Dale \& Uson~(2000), respectively.  The
data for Abell~1413, Abell~2218 and Abell~2670 were obtained and processed
using essentially the same approach.  This section will present only new
data---those for Abell~1413, Abell~2218 and Abell~2670---whereas the data from
all five clusters will be included in the discussion.

\subsection {Optical Spectroscopy}

Long-slit spectroscopy was carried out at the Palomar Observatory 5~m
telescope (Table~\ref{tab:spec_runs}).  The red camera of the Double
Spectrograph (Oke and Gunn~1982) was used to observe the \hal\ (6563~\AA),
\NII\ (6548, 6584 \AA), and \SII\ (6717, 6731~\AA) emission lines.  The spatial
scale of CCD21 (1024$^2$ pixels) was 0$\farcs$468 pixel$^{-1}$.  The
combination of the 1200~lines~mm$^{-1}$ grating and a 2\arcsec\ wide slit
yielded a dispersion of 0.65~\AA~pixel$^{-1}$ and a spectral resolution of
1.7~\AA\ (equivalent to 75\kms\ at 6800~\AA).  Depending on which clusters were
to be observed, the blaze and grating angle were selected according to the
expected redshifted wavelength of \hal.

Deep $I$ and/or $R$ band images (\S~\ref{sec:phot}) were used to select
candidate galaxies as well as to estimate their position angles for follow-up
long-slit spectroscopy.  The spectroscopic observing strategy usually began
with a 10 to 15~minute test exposure on each target.  This allowed an estimate
of the exposure time required in order to adequately sample the outer disk
regions.  Furthermore, the test exposure determined whether the galaxy was even
useful for our purposes: a galaxy may lie in the foreground or background of
the cluster or it may contain little or no \hal\ emission.  If the galaxy was
deemed useful, a sequence of follow-up exposures were taken, with each
integration typically 15 to 30~minutes in duration.  Line emission was detected
in 31 of the 78~galaxies observed in the fields of Abell~1413, Abell~2218 and
Abell~2670 (the detection rate was 26/52 for Abell~2029 and 13/26 for
Abell~2295).  The galaxies observed are listed in
Table~\ref{tab:galaxy_sample}.

Rotation curves have been extracted as discussed in Dale et al.~(1997, 1998).
The \hal\ emission line was used to map the rotation curve except in the case
of the galaxy AGC~630625 (in Abell~2670) where the emission of the \NII\ line
(6584~\AA) extends to a larger distance than that of the \hal\ emission
(``AGC'' refers to the Arecibo General Catalog maintained by M.~P.~Haynes and
R.~Giovanelli).  Small portions of the \hal\ rotation curve of one galaxy
(AGC~630571, again in Abell~2670) have been determined using data from the
\NII\ rotation curve in order to provide information on the shape of the inner
parts and to ensure a consistent estimate of the velocity width.

The rotation curves vary in physical extent, and more importantly, they do not
all reach the optical radius, \Ropt, the distance along the major axis to the
isophote containing 83\% of the $I$~band flux.  Extrapolations to the rotation
curves, and hence adjustments to the velocity width, are made when the \hal\
emission does not extend to \Ropt.  The resulting correction depends on the
shape of the rotation curve and exceeded 4\% for five galaxies: AGC~261226
($\approx 15$\%) in Abell~2218, and AGC~630588 ($\approx 22$\%); AGC~630593
($\approx 19$\%); AGC~630615 ($\approx 10$\%) and AGC~630625 ($\approx 6$\%) in
Abell~2670.  To recover the actual rotational velocity widths, a few more
corrections are necessary.  The additional corrections account for disk
inclination, cosmological broadening, and the smearing of the velocity profile
in the 2\arcsec\ slit width (Dale~1998).

\begin{figure}[ht]
\centerline{\psfig{figure=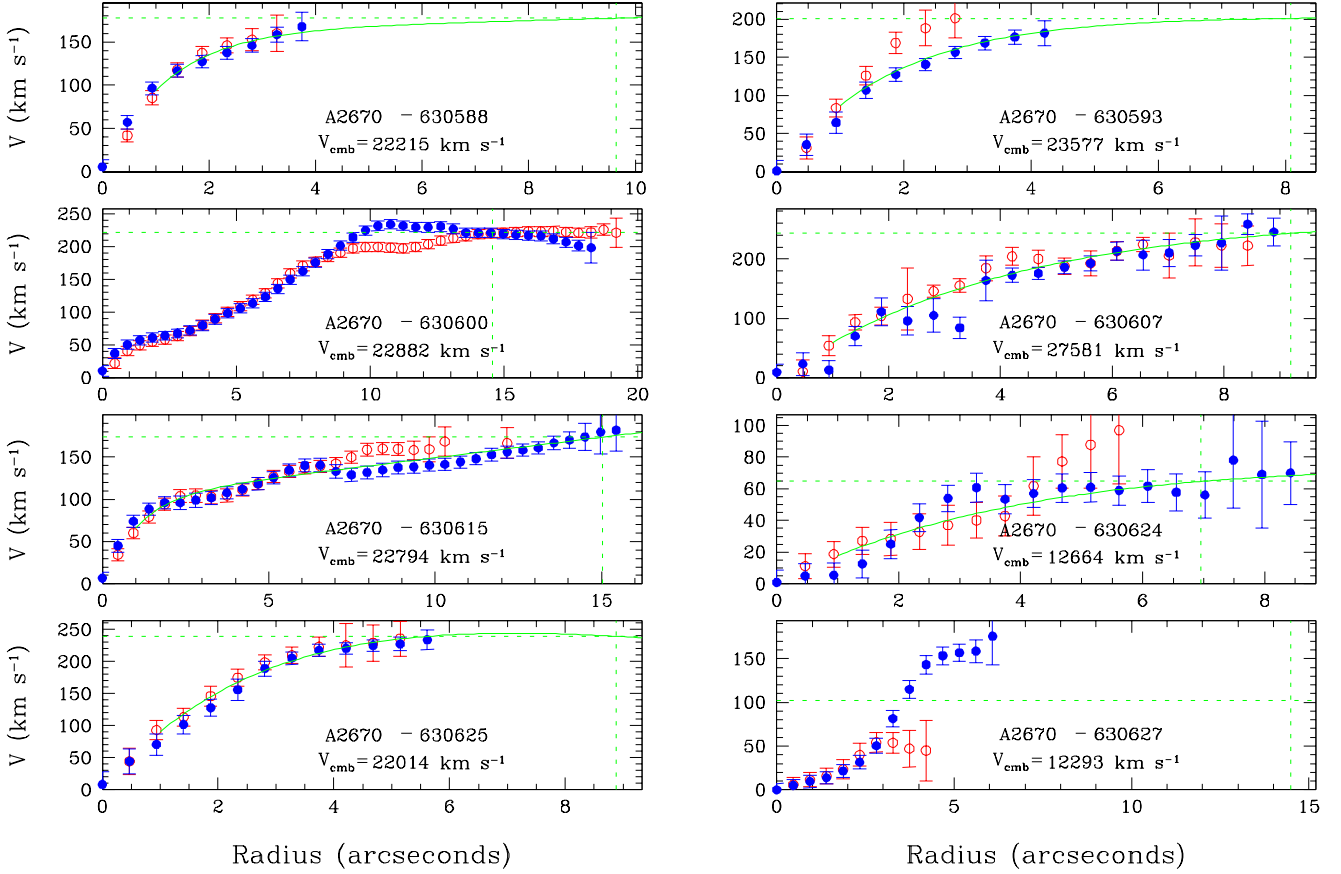,width=7.0in,bbllx=67pt,bblly=348pt,bburx=515pt,bbury=714pt}}
\caption[] {The folded rotation curves.  The error bars include both the uncertainty in the wavelength calibration and the rotation curve fitting routine used.  Names of the galaxy are given along with the CMB radial velocities.  Two dashed lines are drawn: the horizontal line indicates the adopted half velocity width, which in some cases arises from an extrapolation to the rotation curve; the vertical line is at \Ropt, the radius containing 83\% of the $I$ band flux.  A fit to the rotation curve is indicated by a solid line.  Note that the rotation curves are {\it not} deprojected to an edge-on orientation.}
\label{fig:RCs}
\end{figure}

Figure~\ref{fig:RCs} is a display of the rotation curves observed in the fields
of Abell~2218 and Abell~2670.
Notice that the data are highly correlated due to seeing
and guiding jitter.  This is properly taken into account by the fitting
routines (see Dale et al.~1997 and references therein for details).
Table~\ref{tab:spec} below contains the complete set of spectroscopic data
for the galaxies for which useful rotation curves were obtained.  Detailed
comments for some galaxies follow:
  
\small
\noindent A2218-261226: Large rotation curve extrapolation.\\
\noindent A2218-261244: Foreground galaxy.\\
\noindent A2218-261712: Foreground galaxy.\\
\noindent A2218-261715: Foreground galaxy.\\
\noindent A2670-630571: \NII\ patch over inner 3\arcsec; weak line emission;
uncertain disk ellipticity; unfit for Tully-Fisher use.\\
\noindent A2670-630588: Large rotation curve extrapolation.\\
\noindent A2670-630593: Large rotation curve extrapolation; possible flux
contamination from galaxy 42\arcsec\ to the NNW; merger?\\
\noindent A2670-630607: Background galaxy.\\
\noindent A2670-630624: Foreground galaxy; note low~$i$.\\
\noindent A2670-630625: Large rotation curve extrapolation.\\
\noindent A2670-630627: Foreground galaxy; center of light used for rotation
curve spatial and kinematic center.\\
\normalsize

\subsection {Optical Imaging}
\label{sec:phot}

$I$~band photometry of Abell~2218 was obtained for a different project by one
of us (JMU) in collaboration with S. P. Boughn (Haverford) with the 0.9~m
telescope on Kitt Peak National Observatory on 1998~April~20.  They used the
T2KA camera mounted at the f:7.5 Cassegrain focus which resulted in square
pixels, 0$\farcs$68 on a side.  The seeing was good, between 1$\farcs$1
and 1$\farcs$4, which resulted in an effective seeing of $\sim1\farcs$5 due
to the available pixel size.
Two sets of nine partially overlapping frames plus an extra frame at the center
were used to form a mosaic of about 35$\arcmin$ (RA) by 36$\arcmin$ (Dec).  The
mosaic has overlaps of about 3/4 of a frame between immediately adjacent
frames.  All frames were obtained with air masses between 1.21 and 1.25.  The
exposures lasted five minutes.  The data were processed as discussed in Uson,
Boughn and Kuhn (1991, hereafter UBK91).  All frames were used to generate a
``sky-flat'' gain calibration frame.  Since the cluster contains a diffuse halo
that surrounds the central galaxy, an 8$\arcmin$ by 8$\arcmin$ area centered
on the cluster was blanked on all the frames before using them to generate the
sky flat as discussed in UBK91.  Because the atmospheric extinction was low
(the secant-law extinction had a slope of 0.06 mag/airmass) and the frames were
obtained at an approximately constant airmass, no differential atmospheric
correction was applied to the frames.  Absolute calibration was done using
stars from Landolt's $UBVRI$ secondary calibration list (Landolt~1983).

We observed  Abell~2670 in the $I$~band using the Palomar 1.5~m telescope on
the night of September 19, 2001.  We used the $2048^2$ CCD13 camera mounted at
the f:7.5 Cassegrain focus which resulted in square pixels, 0$\farcs$378 on a
side.  The seeing was good, about 1$\farcs$1, which resulted in an effective
seeing of $\sim1\farcs$2. We obtained two sets of nine partially-overlapping
frames in a $3 \times 3$~square pattern plus two sets of five frames in a
``plus sign'' pattern (with half of the spatial offsets used in the previous
arrangement), all of them centered on the cD galaxy.  Secant law calibration
was deduced from stars located on the cluster frames
($0.055 \pm 0.008$ mag/airmass) and agreed with that determined from secondary
Landolt standards.  The observations were made at airmasses between 1.4 and
2.5.  The data were processed as described above, again cutting from each frame
an 8$\arcmin$ by 8$\arcmin$ area centered on the cluster in order to obtained
a ``sky-flat'' gain calibration frame.  Details will be given elsewhere.

$R$~band photometry of Abell~1413 was obtained for a different project by one
of us (JMU) in collaboration with S. P. Boughn with the 0.9~m telescope on Kitt
Peak National Observatory on 1998~April~17.  They used the same setup as for
the $I$~band observations of Abell~2218 discussed above.  The seeing was good,
between 0$\farcs$9 and 1$\farcs$3, which resulted in an effective seeing of
$\sim1\farcs$4.  Three sets of nine partially overlapping frames were used to
form a mosaic of about 35$\arcmin$ by 35$\arcmin$.  The mosaic has overlaps of
about 3/4 of a frame between immediately adjacent frames.  All frames were
obtained with air masses between 1.01 and 1.10 with exposures of 2.5~minutes.
The data were processed in the same way as the data for Abell~2029, discussed
in UBK91.  Because the atmospheric extinction was reasonably low (the
secant-law extinction had a slope of 0.16 mag/airmass) and the frames were
obtained at an approximately constant airmass, no differential atmospheric
correction was applied to the frames.  Absolute calibration was done using five
stars from Landolt's $UBVRI$ secondary calibration list (Landolt~1983) which
were observed four times each at airmasses between 1.12 and 1.16.  The mosaic
was used to select candidate galaxies for the spectroscopic observations.
Because the weather prevented us from obtaining any adequate rotation curves
for this cluster, the absence of $I$~band photometry for this cluster is,
sadly, not a problem at this time.

Flux estimation follows from the data reduction methods discussed in Dale et
al.~(1997, 1998) using both standard and customized IRAF packages.
Cosmological $k$-corrections are applied according to Poggianti~(1997).  The
relevant photometric data are listed in Table~\ref{tab:phot} with the first
column matching that of Table~\ref{tab:spec}.

\section {Results}

\begin{figure}[ht]
\centerline{\psfig{figure=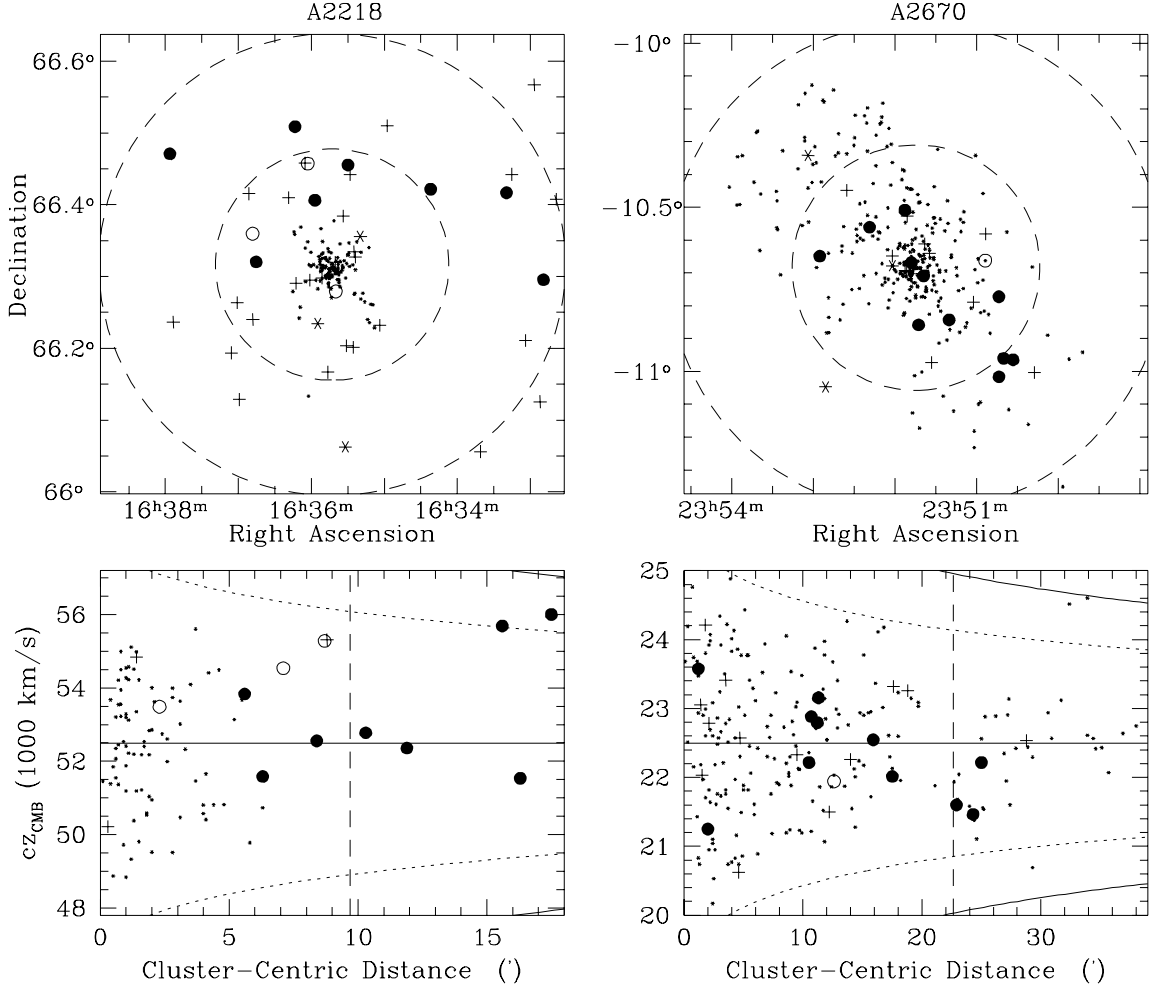,width=6.0in,bbllx=53pt,bblly=164pt,bburx=574pt,bbury=616pt}}
\caption[Cluster Membership] {Sky and velocity distribution of galaxies in the fields of Abell~2218 and Abell~2670.  Circles represent cluster members with measured photometry and widths; if unfilled, widths are poorly determined.  Asterisks identify foreground and background galaxies, and dots give the location of galaxies with known redshift.  Crosses indicate the positions of galaxies lacking emission lines (their redshifts are drawn from the literature).  The dashed lines indicate 1 and 2~Abell radii.  The dotted and solid lines respectively indicate 2~and 3$\sigma$ cluster membership contours (see Equations~8--11 in Carlberg et al. 1997, with $H_o=70$\kms~Mpc$^{-1}$, $\Omega_o=0.3$ and $\Omega_\lambda=0.7$).}
\label{fig:field}
\end{figure}

Plots of the sky distribution of the galaxies observed in Abell~2218 and
Abell~2670 are displayed in the upper panels of Figure~\ref{fig:field}.  The
lower panels present galaxy redshifts versus their projected distances from the
cluster centers.  Combining our observations with those reported in the
literature yields large redshift samples: $N_z=93$ and $N_z=235$ for
Abell~2218 and Abell~2670, respectively.  For Abell~2218 the mean CMB cluster
redshift is c$z_{\rm cmb}=52,497^{+128}_{-170}$\kms\ and the (rest frame) velocity
dispersion $1437\pm100$\kms; for Abell~2670 the numbers are
c$z_{\rm cmb}=22,494^{+87}_{-29}$\kms\ and $\sigma_{\rm rest}=895\pm40$\kms.  Projected
cluster membership contours are derived using the results from the CNOC survey
of galaxy clusters (Carlberg et al.~1997), scaled by the cluster velocity
dispersions.  The 3$\sigma$ contour is indicated by the two solid curves in
each lower panel; the dotted curves show the 2$\sigma$ contours (see, for
example, Balogh et al.~1999).

\subsection {The Distribution of Emission-Line Galaxies}
\label{sec:distribution}
Only 45\% (70 of 156) of the targeted galaxies were detected in \hal.  A small
subset of the non-detected galaxies might lie outside the 30,000\kms\ c$z$
ranges probed (Table~\ref{tab:cluster_sample}).  In contrast, Dale et
al.~(1999) detected line emission in 65\% (582 of 897) of their
$0.02\lesssim z \lesssim 0.06$ targets, using a nearly identical observational
approach.  
This relatively low success rate is likely due to the higher density
of the clusters discussed in this paper, although it might be due in part to
target selection effects.  Whereas Dale et al.~(1999) almost exclusively
targeted late-type cluster galaxies, it is more difficult to avoid targeting
early-type galaxies in more distant clusters.  Balogh et al.~(2002), for
example, profited from $Hubble~Space~Telescope$ imaging to differentiate
between early-types and late-type galaxies in Abell~1689 at $z=0.18$: they find
that more than 90\% of the cluster spirals show \hal\ emission, whereas less
than 10\% of the early-type galaxies are ``securely detected'' in \hal.

\begin{figure}[ht]
\centerline{\psfig{figure=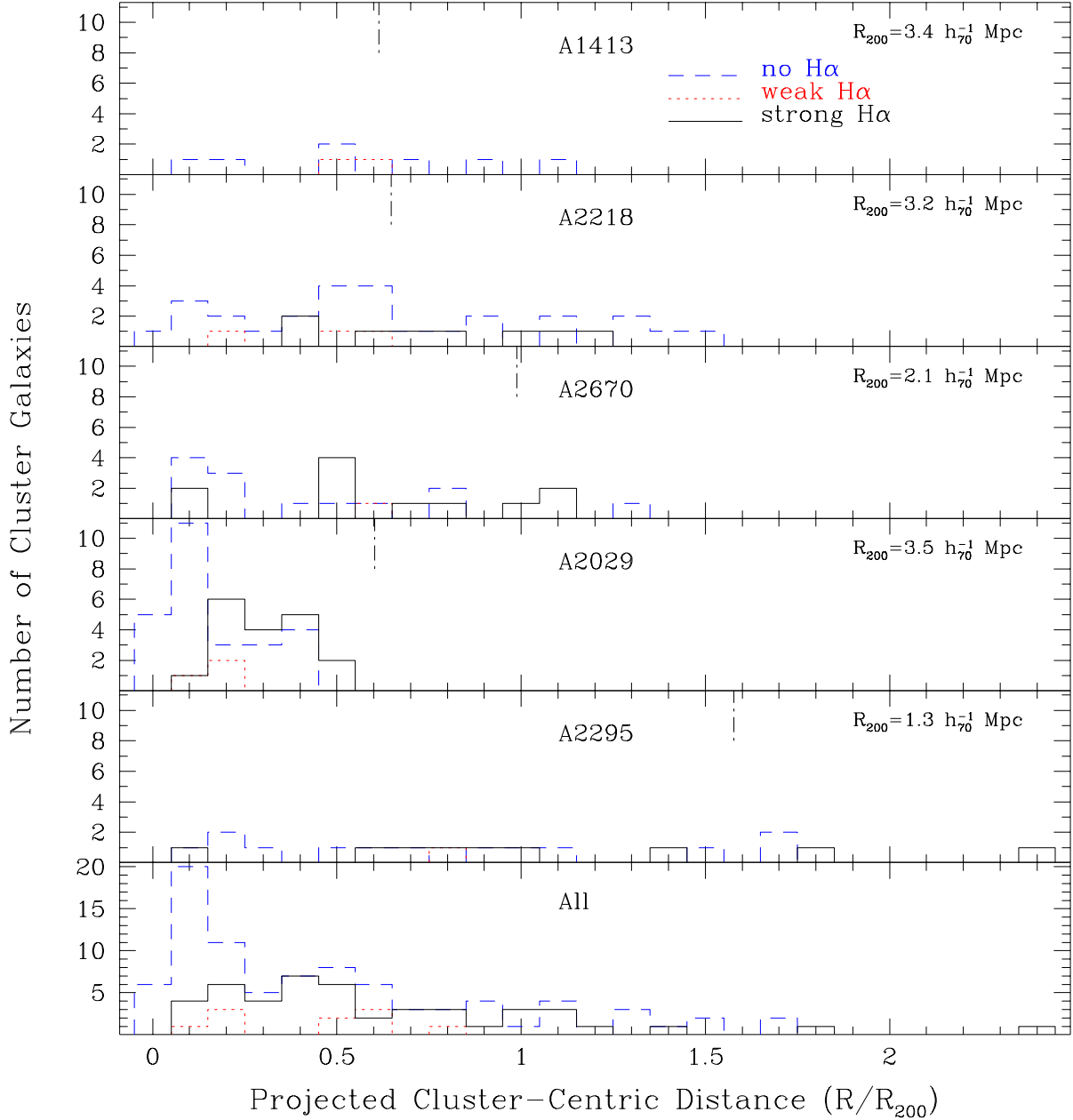,width=6.0in,bbllx=46pt,bblly=100pt,bburx=564pt,bbury=750pt}} 
\caption[Cluster Membership] {The distribution of observed cluster galaxies in Abell~1413, Abell~2029, Abell~2218, Abell~2295, and Abell~2670 (panels 1 through 5) as well as for the combined sample, as a function of the projected distance to the cluster center (in units of $R_{200}$), separated according to the strength of the observed \hal\ line emission.  One Abell radius is indicated by a dot-dash vertical line.  
}
\label{fig:detections}
\end{figure}

Figure~\ref{fig:detections} shows the distribution of the observed galaxies as
a function of projected distance to their respective cluster centers.  Because
the clusters are of different richness, we have scaled the distances using
$R_{200}$, the cluster radius where the mean interior density is 200 times the
critical density, since that radius generally contains the virialized mass
(e.g. Equation~8 in Carlberg et al. 1997).  As pictured in
Figure~\ref{fig:detections}, there is a difference in the spatial distribution
of galaxies with and without optical emission lines (the
search was limited to \hal, \NII, and \SII; see Table~\ref{tab:galaxy_sample}).
The galaxies with no line emission are more concentrated towards the
cluster centers, at a median projected distance of 0.36~$R_{200}$, whereas the
galaxies with line emission are found to lie further out, at a median
projected distance at 0.47~$R_{200}$.  This result is similar to
findings that the star formation rate in cluster galaxies increases for larger
cluster-centric distances (Balogh et al. 1998; Lewis et al.~2002).  Inspection of Figure~\ref{fig:env} shows that our sample selection is skewed towards late-type spirals near the cluster cores, suggesting that the observed distribution of emission-line galaxies is not entirely due to cluster morphological segregation (e.g. Dressler et al.~1997), consistent with the findings of Balogh et al. (1998) and Lewis et al. (2002).  Furthermore,
the panels for the individual clusters show significant segregation in A2029,
A2218 and especially in A1413 where no galaxies showed strong \hal\ emission.  Conversely, no segregation is apparent in the poorer clusters A2295 and A2670
(which although it contains a cD galaxy seems far from dynamical relaxation as
discussed above).

\subsection {Rotation Curve Asymmetry, Shape, and Extent}

\subsubsection {Rotation Curve Asymmetry}

Galaxy-galaxy or galaxy-cluster interactions can disturb disk velocity fields
(e.g. Conselice \& Gallagher~1999), but since they are expected to regularize
within a few rotation cycles, it is likely that rotation curve asymmetries
reflect only the most recent interaction history (see Dale et al.~2001 and
references therein).  As much as half of all galaxies, in clusters and in the
field, show significant rotation curve asymmetries or lopsided \HI\ profiles
(e.g. Richter \& Sancisi~1994; Haynes et al.~1998; Swaters et al.~1999).  

To measure the global rotation curve asymmetry the total area between the
kinematically-folded approaching and receding halves is normalized by the
average area under the rotation curve:
\be
{\rm Asymmetry}={\sum {\vert \mid V(R)| - |V(-R) \vert \mid\over \sqrt{\sigma^2(R) + \sigma^2(-R)}} \over \case{1}{2} \sum {|V(R)|+|V(-R)| \over \sqrt{\sigma^2(R) + \sigma^2(-R)}}}
\label{eq:asymmetry}
\ee
where $\sigma(R)$ is the uncertainty in the data at radial position $R$.  The
sample preferentially includes inclined disk systems ($i\gtrsim45^\circ$),
implying that the asymmetry parameter is more sensitive to noncircular than
nonplanar motions (Kornreich et al.~2000).  The average asymmetry for
$z\sim0.1$ cluster galaxies, $12.6\pm1.2$\%, is similar to that for the sample
of more than 400~cluster galaxies and more than 70~field galaxies at lower
redshift ($14.0\pm0.4$\% and $12.5\pm1.0$\%, respectively) studied by Dale et
al. (1997).  There are no statistically significant differences with (spiral)
galaxy morphological type or cluster-centric distance (see
Figure~\ref{fig:env} and Table~\ref{tab:results}).

\subsubsection {Rotation Curve Shape}

The shape of galaxy rotation curves has been suggested to be a probe of
environmental influences (e.g. Whitmore, Forbes \& Rubin 1988) as the cluster
environment may inhibit dark matter halo formation and
galaxy-galaxy/galaxy-cluster interactions should lead to stripping of part of
the halo mass.  Rotation curve shape is well-studied for galaxies in nearby
clusters (see Dale et al.~2001 and references therein). 

An indication of the rotation curve shape in the outer disk region is the
``outer gradient'' parameter ($OG$) defined in Dale et al.~(2001) as
\be
OG(\%) = 100 \times {V(R_{\rm opt}) - V(0.5R_{\rm opt}) \over V(R_{\rm opt})}
\label{eq:outer_gradient}
\ee  
where $V=V_{\rm rot}\sin i$ is the projected rotational velocity of the disk
at inclination $i$.  Figure \ref{fig:env} displays this parameter as a
function of projected cluster-centric distance.  The average outer gradient
for all $z\sim0.1$ cluster galaxies is $\langle OG \rangle=11.1 \pm 1.2$\%;
a similar value is found for the low redshift cluster and field galaxies
($10.1 \pm 0.4$\% and $10.6 \pm 1.1$\%).  There are no statistically
significant differences with spiral galaxy morphological type or
cluster-centric distance.

\subsubsection {Rotation Curve Extent}

Another possible indicator of prior environmental influences is the radial
extent of the line emission in spiral disks.  In contrast to the inner-galaxy
matter, the more diffuse and peripheral mass is less gravitationally bound and
is more easily stripped.  For example, \HI-deficiency is common for
inner-cluster galaxies (e.g. Solanes et al.~2001), and the outer disk star
formation rate is clearly truncated in Virgo cluster galaxies (Koopmann \&
Kenney~2002).

To extract rotational velocities reliably, good spectroscopic data are needed
in the outermost portions of the galaxy disks.  Therefore, care was taken to
obtain sufficiently deep  integrations to ensure high signal-to-noise data for
the outer portions of the rotation curves.  For the lower redshift cluster
sample of Dale et al.~(2001), the average radial extent of the rotation curves,
\Rcut, for all cluster galaxies is
$\langle R_{\rm cut}/R_{\rm opt} \rangle=1.11 \pm 0.02$; similar values are
found for early-type spirals, late-type spirals, and foreground and background
spirals.  For the $z\sim0.1$ cluster sample, the mean ratio is
$\langle R_{\rm cut}/R_{\rm opt} \rangle=1.18 \pm 0.07$.  Does the extent of
measured emission depend on environment?  Rubin, Waterman \& Kenney~(1999)
found that to be the case for Virgo cluster spirals, though no quantitative
result was given.  Dale et al.~(2001) find a mild trend, a $4.1\pm2.3$\%
increase per\h~Mpc ($2.9\pm1.6$\% increase per\ha~Mpc) of projected
galaxy-cluster core separation.  We have recomputed this trend using
$R_{200}$ for the clusters in their sample and find a more significant increase
in the \hal\ extent of $3.5\pm0.5$\% per $R_{200}$.  The third panel of
Figure~\ref{fig:env} shows the radial extent of the rotation curves versus
cluster-centric distance.  We find no clear trend, an increase of $5\pm14$\%
per $R_{200}$ ($\sim13\pm7$\% per\ha~Mpc).  However, as discussed in
Section~\ref{sec:distribution} above, it seems
that a large fraction of the galaxies located near the centers of these
clusters shows no emission lines, suggesting that passage through the cores
might have stripped the galaxies of most of their gas rather than just trimming
its spatial extent.

\subsubsection {\hal\ Equivalent Width}
The 2dF cluster data confirmed that \hal\ equivalent width (a proxy for the
star formation rate) decreases with decreasing projected cluster-centric
distance, and is not entirely due to morphological segregation in
clusters (Lewis et al.~2002).  Gavazzi et al. (2002) focused on late-types in
an \hal\ study of 369 Virgo and Coma/A1367 galaxies, and found this trend only
occurred for the brighter subset in Virgo.  No firm trend is seen in our data (fourth panel in Figure~\ref{fig:env}).

\subsection {The Tully-Fisher Relation at $z\sim0.1$}
\label{sec:TF_relation}

Dale \& Uson~(2000) made an initial study of the Tully-Fisher relation at
$z\sim0.1$ via deep optical imaging and spectroscopy of Abell~2029.  Based on
a sample of 14 cluster galaxies, the study hinted at a larger intrinsic
Tully-Fisher scatter and a zero point equivalent to that found in nearby
clusters.  The current sample more comprehensively probes the $z\sim0.1$
epoch---43 galaxies spread throughout four different cluster environments.

\begin{figure}[ht]
\centerline{\psfig{figure=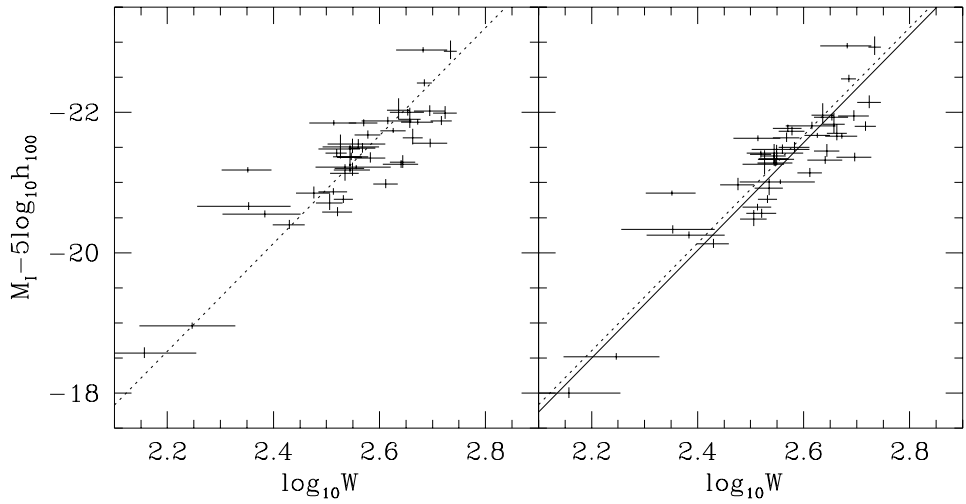,width=5.0in,bbllx=18pt,bblly=408pt,bburx=572pt,bbury=698pt}}
\caption[Cluster Membership] {The Tully-Fisher data for Abell~2029, Abell~2218, Abell~2295, and Abell~2670.  The morphological offsets advocated by Giovanelli et al. (1997) and Dale et al.~(1999)  have been applied to the early-type spirals: $\Delta m_T=-0.1$ mag for the Sb galaxies, and $\Delta m_T=-0.32$ mag for earlier type spirals.  The data in the righthand panel are additionally corrected for cluster peculiar motion and population incompleteness bias.  The dashed line in each panel is the template relation for $0.02\lesssim z \lesssim 0.06$ clusters (Equation~\ref{eq:TFlowz}) obtained by Dale et al. (1999), whereas the solid line in the righthand panel (Equation~\ref{eq:TFhighz}) reflects the average zero point for the $z\sim0.1$ clusters: Abell~2029, Abell~2295, and Abell~2670.}
\label{fig:TF}
\end{figure}

The Tully-Fisher data for all  members of the $z\sim0.1$ clusters are presented
in Figure~\ref{fig:TF} (residuals as a function of cluster environment are
portrayed in the bottom panel of Figure~\ref{fig:env}).  The morphological type offsets for early-type disk
galaxies advocated by Giovanelli et al.~(1997) and Dale et al.~(1999) are
applied: $\Delta m_T=-0.1$ mag for Sb types and $\Delta m_T=-0.32$ mag for
types earlier than Sb.  Included in the plot is the template relation obtained
from the Dale et al.~(1999) all-sky of sample 52~galaxy clusters at
$0.02\lesssim z \lesssim 0.06$:

\be
y = -7.68[\pm0.10]x - 20.905[\pm0.020]~{\rm mag}
\label{eq:TFlowz}
\ee
where $y$ is $M_I - 5\log h_{100}$ and $x$ is log$W_{\rm cor} - 2.5$.  The
uncertainty in the zero point is a combination of the standard statistical
uncertainty (0.018 mag) and the ``kinematical uncertainty'' (0.009 mag) due
to the typical peculiar motions of clusters (see \S~3.1 of Dale et al.~1999).
As explained below, the slope and its uncertainty were adopted from Giovanelli
et al.~(1997).

The data in the righthand panel of Figure~\ref{fig:TF} are additionally
corrected for cluster population incompleteness bias.  In short, because
magnitude-limited imaging preferentially samples the brighter end of the
cluster luminosity function, the observed Tully-Fisher slope is artificially
too shallow and the zero point overluminous.  To remove this effect on the
zero point, Monte Carlo simulations have been performed to recover an
effective true ``parent population,'' given the observed sample and the
scatter in the Tully-Fisher relation (Dale et al.~(1999).  In addition, samples
of nearby clusters can better probe the Tully-Fisher slope: a wide dynamic
range (of rotational velocity widths) is more easily observed in a nearby
sample; owing to the relative propinquity of the Giovanelli et al.~(1997)
sample ($z\lesssim0.02$), and because the Giovanelli et al.~(1997)
prescriptions for extracting Tully-Fisher data are consistent with their
approach, Dale et al.~(1999) adopted the (bias corrected) Giovanelli et
al.~(1997) slope.  If this $I$ band slope from nearby clusters is incorporated
into this study, the fully-corrected $z\sim0.1$ Tully-Fisher relation is

\be
y(z\sim0.1) = -7.68[\pm0.10]x - 20.800[\pm0.065]~{\rm mag}.
\label{eq:TFhighz}
\ee

The zero point is based on data from Abell~2029, Abell~2295, and Abell~2670,
and assumes that these three clusters are on average at rest in rest frame of
the cosmic microwave background.  The scatter in the (bias corrected)
Tully-Fisher data for each cluster is listed in Table~\ref{tab:TF_results}.  
Calculated with respect to the $z\sim0.1$ Tully-Fisher zero point, the
inferred peculiar motion for Abell~2218 is large.  Alternatively, assuming 
that Abell~2218 is at rest with respect to the cosmic background radiation 
leads to a slightly fainter zero point in the Tully-Fisher relation at 
$z \sim 0.18$, about $-20.55$~mag.  However, this is only a $2\sigma$
effect, and is thus not significant at this point.

\section {Discussion and Summary}
\label{sec:summary}

The goal of this survey is to investigate the impact that the intracluster
medium has on galaxies residing in a range of cluster environments at
$z\sim 0.1$.  A total of 156~galaxies were observed spectroscopically in five
clusters ranging from rich, X-ray luminous systems to richness class~0 clusters
undetected in X-rays.  
About half of the sample (45\%) was detected in \hal.

We observe no significant trend in the extent of \hal\ emission as a function
of cluster-centric distance.  This is in sharp contrast to what is observed in
the nearby cluster sample, particularly when the projected cluster-centric
distance is expressed in units of $R_{200}$.  We also fail to detect significant
trends in rotation curve asymmetry, $I$ band Tully-Fisher residuals, rotation
curve shape, and \hal\ equivalent width.  The relatively small number
of rotation curves obtained for the dense clusters might be the main reason for
our inability to detect statistically significant trends with cluster richness
or location within the clusters.  However, it is also possible that the low \hal\
detection rate is a consequence of ram pressure stripping of the galaxies as
they traverse the cores of the clusters, so that the galaxies that we do
detect in \hal\ would have avoided the cluster cores irrespective of their
projected cluster-centric separation.  Indeed, a VLA study of Abell~2670 found
$\sim 30$ \HI\ sources, but {\it none} within 0.5\ha~Mpc of the cluster center
(van Gorkom~1996).  But a deep VLA survey of Abell~2029, sensitive to a
5$\sigma$ \HI\ mass of $5 \times 10^{8}$~M$_\odot$ within 1.5\ha~Mpc of the
cluster center, has detected only three sources (Uson, in preparation).

\begin{figure}[ht]
\centerline{\psfig{figure=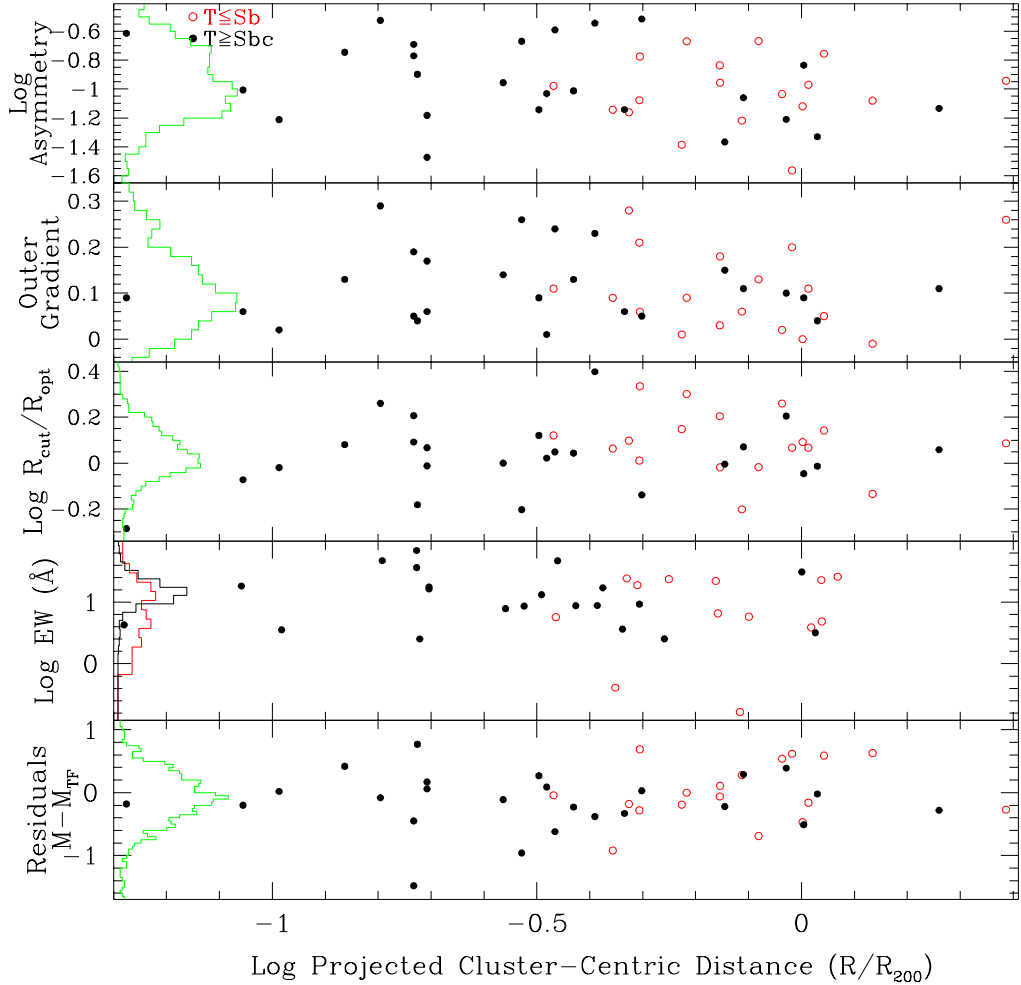,width=5in,bbllx=27pt,bblly=143pt,bburx=565pt,bbury=724pt}}
\caption[Cluster Membership] {{\bf Top panel}:  Rotation curve asymmetry (Equation \ref{eq:asymmetry}).  {\bf Second panel}:  The outer gradient parameter for the rotation curves (Equation \ref{eq:outer_gradient}).  {\bf Third panel}:  The maximum radial extent of the observed emission line within the disk, normalized to the semi-major axis containing 83\% of the $I$ band flux.  The histograms in the top three (and bottom) panels are the distributions for 441 galaxies in 52 lower redshift Abell clusters (Dale et al. 1999).  {\bf Fourth panel}:  \hal\ equivalent width.  Open circles indicate Sb and earlier cluster galaxies while filled circles show Sbc and later cluster galaxies.  The histograms at the left are the \hal\ equivalent widths found by Kennicutt (1998).  The more sharply peaked histogram is for $T\geq$~Sbc, while the other histogram is for $T\leq$~Sb.  {\bf Bottom panel}:  Residuals of the Tully-Fisher data as a function of projected cluster-centric distance (Equation~\ref{eq:TFhighz}).}
\label{fig:env}
\end{figure}

There is a clear difference in the cluster-centric distribution of actively
star-forming and quiescent galaxies, with the star-forming galaxies
preferentially found more in the cluster peripheries.  This result is perhaps
not surprising, given that the well-known cluster morphology-density relation
likely contributes to this effect.  Although ground-based imaging precluded
perfect sample selection, we strived to only observe spiral galaxies.
Moreover, as can be seen in Figure~\ref{fig:env}, our sample selection appears
to have been skewed towards late-type spirals near the cluster cores.  Thus, if
the morphology-density relation has contributed to our observed distribution of
strong and weak \hal\ emitters, our sample selection seems to have minimized
its impact.

The zero-point in the $z\sim0.1$ $I$ band Tully-Fisher relation
(Equation~\ref{eq:TFhighz}) is a bit fainter (by 1.5$\sigma$) than that
observed for nearby clusters (Equation~\ref{eq:TFlowz}).  The data for
Abell~2218, even more distant at $z\sim0.18$, are an additional 1.8$\sigma$
fainter than the $z\sim0.1$ template.  Are these discrepancies related to the
comparatively high redshift of the clusters?  Is this discrepancy evidence of
evolution or observational biases?  It is unclear whether or not the difference
is due to systematic effects in the observational program: though seeing
effects push deprojected rotational velocity widths to artificially high
values, ``slit smearing'' of the disk rotation profile conversely biases the
widths low.  In case the correction recipes applied are unfit at this
relatively high redshift, the simulations and empirical prescriptions used to
generate the correction recipes have been revisited.  For example, Dale et
al.~(1997) have calibrated the effects of seeing on the inferred inclinations.
They found a linear relation between the circularization of isophotes and the
ratio of the seeing full-width half-maximum to face-on disk scale length
$\eta$:
\be
\epsilon_{\rm true}/\epsilon_{\rm obs} =  1 + 0.118 \; \eta, 
\ee
where $\epsilon_{\rm true}$ and $\epsilon_{\rm obs}$ are the ``intrinsic'' and
``observed'' disk ellipticities.  However, they only tested this relation for
galaxies observed out to $z\sim0.06$.  Using a separate technique from that
explored in Dale et al.~(1997), we are able to both reproduce the seeing
correction of Dale et al.~(1997) over their sample's $z\lesssim0.06$ redshift
range, and find that it is applicable even for galaxies as distant as those in
Abell~2218.  In the simulations artificial, inclined disk galaxies (created
using MKOBJECTS in the ARTDATA package of IRAF) are progressively smoothed such
that they exhibit increasingly circular isophotes.  Isophotal ellipses are fit
to the simulated galaxies using the same techniques used for the observed
galaxies.  The relation first quantified by Dale et al.~(1997) tends to
non-linearity only when $\eta$ exceeds 2.0, and all of the galaxies observed in
Abell~2218 have this ratio less than 1.8.

Both the total and intrinsic scatter of the $I$ band Tully-Fisher relation at
$z \sim 0.1$ are similar to that in nearby clusters.  The total scatter is
comparable to that found in Giovanelli et al.~(1997) and Dale et al.~(1999):
0.35 and 0.38 mag, respectively.  The intrinsic scatter in the Tully-Fisher
relation is the portion that cannot be reconciled by measurement uncertainties.
This parameter presumably reflects the various feedback mechanisms and range
of parameter space involved in galaxy formation.  Andersen et al.~(2001)
suggest that perhaps up to half of the intrinsic scatter can be accounted for
if disk galaxies simply have an average inherent ellipticity of 0.05, a value
they find for their sample of seven nearly face-on spiral galaxies.  Compared
to the $\sim0.25$~mag intrinsic scatter found by Dale et al.~(1999) in clusters
at lower redshifts, there is no significant difference in the intrinsic
$I$~band Tully-Fisher scatter in these four clusters at $z\sim0.1$.  In short,
no significant evolution in the $I$ band Tully-Fisher properties of cluster
galaxies appears to have occurred since $z \sim 0.1$.

\acknowledgements 
We are grateful for the assistance Richard Cool provided in preliminary
explorations for correlations with cluster X-ray data.  We thank Steve
Boughn for his permission to use the unpublished
{\it I}-band observations of Abell~2029 and Abell~2218 and Heinz Andernach
for his insight on the redshifts of Abell~2218.  The comments of the referee helped to improve the presentation of this work.  The results presented in this
paper are based on observations carried out at the Palomar Observatory (PO) and
at the Kitt Peak national Observatory (KPNO).  The Hale Telescope at the PO is
operated by the California Institute of Technology under a cooperative
agreement with Cornell University and the Jet Propulsion Laboratory.  KPNO is
operated by the Association of Universities for Research in Astronomy, Inc.,
under a cooperative agreement with the National Science Foundation.  This
research has made use of the NASA/IPAC Extragalactic Database (NED) which is
operated by the Jet Propulsion Laboratory, California Institute of Technology,
under contract with NASA, and the Image Reduction and Analysis Facility (IRAF)
which is distributed by the National Optical Astronomy Observatories, which are
operated by the Association of Universities for Research in Astronomy, Inc.,
under a cooperative agreement with the National Science Foundation.  The NRAO
is a facility of the National Science Foundation which is operated under
cooperative agreement by Associated Universities, Inc.  The Digitized Sky Surveys were produced at the Space Telescope Institute under U.S. Government grant NAG W-2166.  The images of these surveys are based on photographic data obtained using the Oschin Schmidt Telescope on Palomar Mountain and the UK Schmidt Telescope.

\scriptsize
\begin {deluxetable}{ccccccclccc}
\def\a{\tablenotemark{a}}
\def\b{\tablenotemark{b}}
\tablewidth{510pt}
\tablecaption{The $z\sim0.1$ Cluster Sample}
\tablehead{
\colhead{Cluster} & \colhead{R.A.} & \colhead{Decl.} & \colhead{$l,b$} & \colhead{c$z_\odot$} & \colhead{c$z_{\rm cmb}$} & \colhead{$\sigma_{\rm 1d,rest}$} & \colhead{$N_z$} & \colhead{$L_X^a$} & \colhead{$R^b$}
\\
\colhead{} & \colhead{(B1950)} & \colhead{(B1950)} & \colhead{(\degr)} & \colhead{(\kmsb)} & \colhead{(\kmsb)} & \colhead{(\kmsb)} & \colhead{} & \colhead{($10^{37}$~W)} & \colhead{}
}
\startdata
A1413 & 115248 & $+$233900 & 226,$+$77 & 42453\p570 & 42767 & 1490 & ~~9 & ~10.3\p0.5 & 4.5\b  \\
A2029 & 150830 & $+$055700 & ~~7,$+$51 & 23220\p170 & 23400 & 1471 & ~92 & ~16.4\p0.6 & 4.4\b  \\
A2218 & 163542 & $+$661900 & ~98,$+$38 & 52514\p170 & 52497 & 1437 & ~93 & ~~9.3\p0.4 & 4.3\b  \\
A2295 & 180018 & $+$691300 & ~99,$+$30 & 24623\p200 & 24555 & ~563 & ~~9 & $<$0.1     & 0  \\
A2670 & 235136 & $-$104100 & ~81,$-$69 & 22841\p~60 & 22494 & ~895 & 235 & ~~2.0\p0.1 & 1.4\b  \\
\enddata
\tablenotetext{a}{Measured over the range 0.5--4.5~keV and covering a 0.7\ha~Mpc radius by Jones \& Forman (1999).}
\tablenotetext{b}{``Abell richness class'' re-estimated by Dressler (1978) from the redshift of these clusters.  Abell's original estimates were $R=3$  (A1413), 2 (A2029), 4 (A2218) and 3 (A2670).}
\label{tab:cluster_sample}
\end{deluxetable}
\normalsize

\scriptsize
\begin {deluxetable}{llcccccc}
\def\a{\tablenotemark{a}}
\def\b{\tablenotemark{b}}
\tablewidth{510pt}
\tablecaption{Palomar 5~m Spectroscopy Observing Runs\a}
\tablehead{
\colhead{Date} & \colhead{Clusters} & \colhead{Nights} & \colhead{$N_{\rm RC}$} & \colhead{\hal} & \colhead{Approx.} & \colhead{Wavelength} & \colhead{c$z$ Range}
\\
\colhead{} & \colhead{Observed} & \colhead{Used} & \colhead{} & \colhead{Rate} & \colhead{Seeing} & \colhead{Coverage} & \colhead{(\hal)}
\\
\colhead{} & \colhead{} & \colhead{} & \colhead{} & \colhead{} & \colhead{(\arcsec)} & \colhead{(\AA)} & \colhead{(\kmsb)}
}
\startdata
1999 Apr 20-21    & A2218       & 0.7\b& ~1  & ~1/~8 & 1.5--2~~ & 7250--7920 & 30,300--62,100\\
1999 May 14       & A2218       & 1    & ~3  & ~4/11 & 1.5--2.5 & 7330--8000 & 33,500--65,700\\
1999 Oct 31-Nov 1 & A2670       & 2\b  & 16  & 15/28 & 1.5--2~~ & 6780--7450 & ~9,800--40,300\\
2001 May 20-21    & A1413,A2218 & 2    & ~7  & 12/31 & 1.5--2.5 & 7330--8000 & 33,500--65,700\\
\enddata
\tablenotetext{a}{The data for Abell~2029 and Abell~2295 are described in Dale \& Uson (2000) and Dale et al. (1998), respectively.}
\tablenotetext{b}{A portion of the observing run was devoted to another project.}
\label{tab:spec_runs}
\end{deluxetable}
\normalsize


\scriptsize
\begin {deluxetable}{lcccc|lcccc}
\tablewidth{550pt}
\tablecaption{Target Galaxy Sample}
\tablehead{
\colhead{Galaxy} & \colhead{R.A.} & \colhead{Decl.} & \colhead{c$z_{\odot}$} & \colhead{RC} & \colhead{Galaxy} & \colhead{R.A.} & \colhead{Decl.} & \colhead{c$z_{\odot}$} & \colhead{RC}
\\
\colhead{} & \colhead{(B1950)} & \colhead{(B1950)} & \colhead{(\kmsb)} & \colhead{} & \colhead{} & \colhead{(B1950)} & \colhead{(B1950)} & \colhead{(\kmsb)} & \colhead{}
}
\startdata

A1413-211894 & 115208.1 & $+$233828 &  ...  & 0 & A2218-261717 & 163612.6 & $+$663031 & 52378 & 1 \\
A1413-211903 & 115216.8 & $+$235549 &  ...  & 0 & A2218-261274 & 163618.1 & $+$662436 &  ...  & 0 \\
A1413-211916 & 115228.0 & $+$235115 &  ...  & 0 & A2218-261718 & 163644.7 & $+$661913 & 51601 & 1 \\
A1413-211924 & 115238.6 & $+$235007 & 42900 & 2 & A2218-261719 & 163647.4 & $+$661425 &  ...  & 0 \\
A1413-211934 & 115243.6 & $+$234059 & 42844 & 0 & A2218-261720 & 163647.5 & $+$662135 & 54557 & 2 \\
A1413-211938 & 115244.8 & $+$234332 &  ...  & 0 & A2218-261286 & 163650.7 & $+$662455 &  ...  & 0 \\
A1413-211956 & 115301.1 & $+$234748 &  ...  & 0 & A2218-261721 & 163658.6 & $+$660743 &  ...  & 0 \\
A1413-211957 & 115301.6 & $+$234725 & 41400 & 2 & A2218-261289 & 163700.4 & $+$661548 &  ...  & 0 \\
A1413-211968 & 115331.3 & $+$235648 &  ...  & 0 & A2218-261290 & 163705.2 & $+$661137 &  ...  & 0 \\
A2218-261222 & 163237.5 & $+$662428 &  ...  & 0 & A2218-261723 & 163752.9 & $+$661412 &  ...  & 0 \\
A2218-261223 & 163248.0 & $+$661744 & 56018 & 1 & A2218-261291 & 163755.8 & $+$662815 & 51553 & 1 \\
A2218-261702 & 163250.8 & $+$660732 &  ...  & 0 & A2670-630560 & 235008.6 & $-$110012 & 22883 & 0 \\
A2218-261224 & 163255.6 & $+$663400 &  ...  & 0 & A2670-630563 & 235024.5 & $-$105752 & 21810 & 1 \\
A2218-261703 & 163302.7 & $+$661240 &  ...  & 0 & A2670-630565 & 235031.5 & $-$105736 & 21948 & 1 \\
A2218-261704 & 163314.2 & $+$662632 &  ...  & 0 & A2670-630567 & 235034.8 & $-$110058 & 22561 & 1 \\
A2218-261705 & 163318.4 & $+$662500 & 55706 & 1 & A2670-630568 & 235034.9 & $-$104620 & 22892 & 1 \\
A2218-261706 & 163340.0 & $+$660322 &  ...  & 0 & A2670-630570 & 235044.7 & $-$103449 & 22607 & 0 \\
A2218-261226 & 163421.0 & $+$662518 & 52793 & 1 & A2670-630571 & 235044.9 & $-$103944 & 22291 & 2 \\
A2218-261708 & 163456.9 & $+$663036 &  ...  & 0 & A2670-630574 & 235053.7 & $-$104718 & 21843 & 0 \\
A2218-261709 & 163502.8 & $+$661355 &  ...  & 0 & A2670-630577 & 235111.6 & $-$105034 & 23501 & 1 \\
A2218-261244 & 163518.8 & $+$662121 & 42855 & 1 & A2670-630584 & 235124.6 & $-$105820 & 23662 & 0 \\
A2218-261245 & 163523.0 & $+$661938 &  ...  & 0 & A2670-630585 & 235126.2 & $-$103826 & 23757 & 0 \\
A2218-261247 & 163524.0 & $+$662003 &  ...  & 0 & A2670-630318 & 235129.7 & $-$103642 & 20968 & 0 \\
A2218-261710 & 163524.6 & $+$661204 &  ...  & 0 & A2670-630317 & 235130.4 & $-$104231 & 21597 & 1 \\
A2218-261248 & 163527.3 & $+$662630 &  ...  & 0 & A2670-630319 & 235132.2 & $-$104208 & 22380 & 0 \\
A2218-261711 & 163529.1 & $+$662719 & 52572 & 1 & A2670-630588 & 235134.1 & $-$105129 & 22562 & 1 \\
A2218-261251 & 163530.2 & $+$661213 &  ...  & 0 & A2670-630322 & 235139.2 & $-$103951 & 23398 & 0 \\
A2218-261712 & 163531.1 & $+$660345 & 37100 & 2 & A2670-630593 & 235139.5 & $-$104006 & 23924 & 1 \\
A2218-261252 & 163532.9 & $+$662302 &  ...  & 0 & A2670-630324 & 235142.6 & $-$103136 & 22674 & 0 \\
A2218-261713 & 163538.9 & $+$661645 & 53500 & 2 & A2670-630597 & 235143.0 & $-$104138 & 24556 & 0 \\
A2218-261259 & 163539.5 & $+$661912 &  ...  & 0 & A2670-630599 & 235144.0 & $-$104136 & 23128 & 0 \\
A2218-261714 & 163545.7 & $+$661000 &  ...  & 0 & A2670-630600 & 235144.2 & $-$103031 & 23229 & 1 \\
A2218-261265 & 163550.3 & $+$661751 & 54862 & 0 & A2670-630328 & 235153.3 & $-$103855 & 22918 & 0 \\
A2218-261715 & 163554.1 & $+$661404 & 39100 & 2 & A2670-630607 & 235153.3 & $-$104040 & 27928 & 1 \\
A2218-261268 & 163556.3 & $+$662422 & 53852 & 1 & A2670-630615 & 235210.2 & $-$103337 & 23141 & 1 \\
A2218-261269 & 163600.5 & $+$661742 &  ...  & 0 & A2670-630621 & 235226.9 & $-$102654 & 23605 & 0 \\
A2218-261271b& 163602.1 & $+$662726 & 55300 & 2 & A2670-630624 & 235242.7 & $-$110249 & 13010 & 1 \\
A2218-261271 & 163604.4 & $+$662730 &  ...  & 0 & A2670-630625 & 235246.9 & $-$103855 & 22361 & 1 \\
A2218-261272 & 163611.9 & $+$661725 &  ...  & 0 & A2670-630627 & 235255.7 & $-$102030 & 12640 & 2 \\
\enddata
\tablecomments{
Col. 1: Abell cluster and coding number in the Arecibo General Catalog.
Cols. 2 and 3: Coordinates have been obtained from the Digitized Sky Survey catalog and are accurate to $<$ 2\arcsec.
Col. 4: The heliocentric radial velocity.  The redshift measurements of the galaxies without emission lines were obtained from NED.  They have been previously derived by others using absorption-line spectra.
Col. 5: Characterization of the optical emission lines: 0=no lines present; 1=strong throughout much of the disk; 2=weak or nuclear only.}
\label{tab:galaxy_sample}
\end{deluxetable}
\normalsize

\scriptsize
\begin {deluxetable}{crcccc}
\def\a{\tablenotemark{a}}
\def\b{\tablenotemark{b}}
\tablewidth{450pt}
\tablecaption{Galaxy Spectroscopic Parameters}
\tablehead{
\colhead{Galaxy} & \colhead{$T_{\rm exp}$} & \colhead{c$z_{\rm cmb}$} & \colhead{$W_{\rm obs}$} & \colhead{$i$} & \colhead{$W_{\rm cor}$}
\\
\colhead{} & \colhead{(s)} & \colhead{(\kmsb)} & \colhead{(\kmsb)} & \colhead{(\degr)} & \colhead{(\kmsb)}
\\
\colhead{(1)} & \colhead{(2)} & \colhead{(3)} & \colhead{(4)} & \colhead{(5)} & \colhead{(6)}
}
\startdata
A2218-261223  & 3900 & 56003\p07 & 397  & 70 & 370\p21\\
A2218-261705  & 4800 & 55690\p07 & 538  & 60 & 530\p27\\
A2218-261226\a& 4500 & 52777\p10 & 300  & 58 & 354\p28\\
A2218-261244\a&  900 & 42838\p14 & 427  & 69 & 413\p24\\
A2218-261711  & 4800 & 52555\p07 & 426  & 69 & 409\p22\\
A2218-261268  & 5700 & 53835\p10 & 402  & 64 & 382\p26\\
A2218-261717  & 4500 & 52360\p07 & 367  & 65 & 350\p23\\
A2218-261718  & 6300 & 51583\p10 & 305  & 66 & 299\p23\\
A2218-261291  & 3300 & 51534\p08 & 398  & 57 & 441\p25\\
A2670-630563  & 3600 & 21463\p10 & 372  & 53 & 450\p33\\
A2670-630565  &  900 & 21601\p10 & 342  & 66 & 372\p23\\
A2670-630567  & 1800 & 22215\p10 & 528  & 70 & 521\p25\\
A2670-630568  & 2400 & 22545\p10 & 453  & 79 & 454\p22\\
A2670-630571\a& 1200 & 21944\p10 & 131  & 54 & 154\p31\\
A2670-630577  & 2400 & 23154\p10 & 390  & 61 & 422\p23\\
A2670-630317  & 1200 & 21250\p10 & 323  & 63 & 352\p27\\
A2670-630588\a& 2400 & 22215\p10 & 292  & 79 & 354\p43\\
A2670-630593\a&  900 & 23577\p10 & 338  & 71 & 413\p45\\
A2670-630600  & 3600 & 22882\p10 & 443  & 90 & 432\p21\\
A2670-630607\a&  600 & 27581\p10 & 427  & 72 & 486\p26\\
A2670-630615  & 3000 & 22794\p10 & 316  & 64 & 363\p45\\
A2670-630624\a&  300 & 12664\p10 & 123  & 42 & 197\p37\\
A2670-630625\a& 4200 & 22014\p10 & 451  & 70 & 495\p33\\
A2670-630627\a&  600 & 12293\p10 & 204  & 67 & 222\p23\\
\enddata
\tablecomments{
Col. 2: The spectral exposure time.
Col. 3: The recessional velocity of the galaxy in the CMB reference frame, assuming a Sun-CMB relative velocity of 369.5\kms\ towards  $(l,b)=264.4^\circ,48.4^\circ)$ (Kogut et al.~1993).
Col. 4: The observed velocity width.
Col. 5: The adopted inclination $i$ of the plane of the disk to the line-of-sight (90$^\circ$ corresponds to an edge-on perspective); the derivation of $i$ and its associated uncertainty are discussed in \S~4 of Dale et al.~(1997).
Col. 6: The velocity width at \Ropt\ converted to an edge-on perspective, corrected for the shape of the rotation curve, cosmological broadening, and the smearing effects due to the finite width of the slit of the spectrograph.  The uncertainty takes into account both measurement errors and uncertainties arising from the corrections.}
\tablenotetext{a}{See notes on individual objects in \S~\ref{sec:phot}.}
\label{tab:spec}
\end{deluxetable}
\normalsize

\scriptsize
\begin {deluxetable}{crrrccccc}
\def\a{\tablenotemark{a}}
\tablewidth{460pt}
\tablecaption{Galaxy Photometric Parameters}
\tablehead{
\colhead{Galaxy}       & \colhead{T}             & \colhead{$\theta$}  & 
\colhead{PA}           & \colhead{$\epsilon$}    & 
\colhead{$R_{\rm d}$}  & \colhead{$R_{\rm opt}$} & 
\colhead{$m_{I}$}      & \colhead{$M_{I}-5\log h_{100}$}	  
\\
\colhead{}	       & \colhead{}              & \colhead{(\arcmin)} & 
\colhead{(\degr)}      & \colhead{}		 &
\colhead{(\arcsec)}    & \colhead{(\arcsec)}	 & 
\colhead{}             & \colhead{}		   
\\
\colhead{(1)} & \colhead{(2)} & \colhead{(3)} & \colhead{(4)} & \colhead{(5)} &
\colhead{(6)} & \colhead{(7)} & \colhead{(8)} & \colhead{(9)}
}
\startdata
A2218-261223  & 3: &  18 &   7 & 0.613\p0.049 &  2.2 &  6.0 & 17.35 & -21.39\p0.07\\
A2218-261705  & 2: &  16 & 100 & 0.464\p0.033 &  1.4 &  4.4 & 17.06 & -21.67\p0.09\\
A2218-261226\a& 1~ &  10 &  17 & 0.432\p0.043 &  1.3 &  4.0 & 17.75 & -20.86\p0.10\\
A2218-261244\a& 2~ &   3 &  18 & 0.596\p0.064 &  4.2 &  8.5 & 16.32 & -21.84\p0.10\\
A2218-261711  & 3: &   8 &   0 & 0.581\p0.040 &  1.2 &  3.7 & 17.72 & -20.88\p0.06\\
A2218-261268  & 5~ &   6 & 114 & 0.540\p0.063 &  1.6 &  5.8 & 17.25 & -21.35\p0.07\\
A2218-261717  & 3~ &  12 &  45 & 0.535\p0.040 &  1.8 &  5.0 & 17.50 & -21.10\p0.06\\
A2218-261718  & 6~ &   6 & 175 & 0.566\p0.058 &  1.0 &  3.8 & 17.75 & -20.85\p0.09\\
A2218-261291  & 2: &  16 &  67 & 0.427\p0.021 &  1.1 &  3.6 & 17.59 & -20.97\p0.09\\
A2670-630563  & 4~ &  24 & 130 & 0.397\p0.042 &  4.3 & 13.0 & 14.66 & -22.00\p0.05\\
A2670-630565  & 5~ &  23 &  83 & 0.583\p0.036 &  4.0 & 12.0 & 14.82 & -21.85\p0.05\\
A2670-630567  & 3~ &  25 & 135 & 0.565\p0.048 &  4.3 & 14.3 & 14.95 & -21.78\p0.07\\
A2670-630568  & 2~ &  16 & 159 & 0.723\p0.045 &  3.2 &  9.8 & 15.18 & -21.58\p0.14\\
A2670-630571\a& 4~ &  13 &  43 & 0.206\p0.112 &  3.3 & 10.8 & 15.12 & -21.64\p0.07\\
A2670-630577  & 5~ &  11 &  51 & 0.491\p0.025 &  2.6 &  7.7 & 15.02 & -21.74\p0.04\\
A2670-630317  & 5~ &   2 & 105 & 0.622\p0.063 &  2.0 &  7.2 & 15.39 & -21.37\p0.05\\
A2670-630588\a& 5~ &  11 &  97 & 0.760\p0.009 &  2.9 &  9.6 & 15.25 & -21.51\p0.10\\
A2670-630593\a& 7~ &   1 & 110 & 0.643\p0.037 &  2.3 &  8.1 & 14.88 & -21.88\p0.05\\
A2670-630600  & 2: &  11 & 138 & 0.804\p0.015 &  4.7 & 14.6 & 15.05 & -21.71\p0.14\\
A2670-630607\a& 5~ &   4 & 145 & 0.664\p0.018 &  3.2 &  9.2 & 15.50 & -21.70\p0.05\\
A2670-630615  & 3: &  11 & 151 & 0.543\p0.057 &  5.4 & 15.0 & 15.31 & -21.45\p0.07\\
A2670-630624\a& 7~ &  27 & 115 & 0.255\p0.040 &  2.1 &  7.0 & 15.86 & -19.65\p0.09\\
A2670-630625\a& 3~ &  18 &  86 & 0.540\p0.099 &  2.7 &  8.9 & 14.84 & -21.92\p0.08\\
A2670-630627\a& 1: &  28 &  80 & 0.572\p0.043 &  5.7 & 14.5 & 14.65 & -20.80\p0.10\\
\enddata
\tablecomments{
Col. 2: Morphological type code: 1 corresponds to Sa, 3 to Sb, 5 to Sc and so on.  The codes are assigned after visually inspecting the $I$ band images and noting the value of $R_{\rm 75}/R_{\rm 25}$, where $R_X$ is the radius containing X\% of the $I$ band flux.  This ratio is a measure of the central concentration of the flux which was computed for a variety of bulge-to-disk ratios.  Given the limited resolution of the images, some of the inferred types are rather uncertain; uncertain types are followed by a colon.
Col. 3: The cluster-centric distance.
Col. 4: Position angle used for spectrograph slit (North: 0$^{\circ}$, East: 90$^{\circ}$).
Col. 5: Ellipticity of the disk corrected for seeing effects as described in \S~\ref{sec:summary}.
Col. 6: The (exponential) disk scale length.
Col. 7: The distance along the major axis to the isophote containing 83\% of the $I$ band flux.
Col. 8: The measured $I$ band magnitude, extrapolated to 8 disk scale lengths assuming that the surface brightness profile of the disk is well described by an exponential function.  $k$-corrections and allowances for Galactic and internal extinction are made.
Col. 9: The absolute magnitude, computed assuming that the galaxy is at the distance indicated by the cluster redshift if the galaxy is a cluster member within 1$R_{\rm A}$ of the cluster center, or else by the galaxy redshift.  The calculation assumes $H_\circ = 100h_{100}$\kms\ Mpc$^{-1}$.  This parameter is calculated after expressing the redshift in the CMB frame and neglecting any peculiar motion.  The uncertainty on the magnitude is the sum in quadrature of the measurement errors and the estimate of the uncertainty in the corrections applied to the measured parameter.}
\tablenotetext{a}{See notes on individual objects in \S~\ref{sec:phot}.}
\label{tab:phot}
\end{deluxetable}
\normalsize


\scriptsize
\begin {deluxetable}{cccccccc}
\def\a{\tablenotemark{a}}
\def\b{\tablenotemark{b}}
\tablewidth{470pt}
\tablecaption{Results}
\tablehead{
\colhead{Spiral}&
\colhead{\hal } &
\colhead{Median}&
\colhead{Median}&
\colhead{RC}    &
\colhead{RC}    &
\colhead{RC}    &
\colhead{\hal}  
\\
\colhead{Galaxy} &
\colhead{Rate} &                           
\colhead{$R_{\rm H\alpha}$} &
\colhead{$R_{\rm no~H\alpha}$} &  
\colhead{Asymm.} &                
\colhead{Shape} &                    
\colhead{Extent} &                   
\colhead{EW}                    
\\
\colhead{Sample} &
\colhead{} &
\colhead{($R_{200}$)} &
\colhead{($R_{200}$)} &
\colhead{(\%)} &
\colhead{(\%)} &
\colhead{(\Ropt)} &
\colhead{(\AA)} 
}
\startdata
$z\sim0.1$ clusters&~70 of 156& 0.47  & 0.36  &12.6\p1.2&11.1\p1.2&1.18\p0.07&15.9\p2.7  \\
$T<$~Sbc           & \nodata  &\nodata&\nodata&10.9\p1.3&10.4\p2.1&1.29\p0.09&12.7\p2.8  \\
$T>$~Sb            & \nodata  &\nodata&\nodata&13.8\p1.7&11.6\p1.5&1.10\p0.09&17.9\p4.0  \\
\hline
$z<0.1$ clusters   &582 of 897& 0.75  &\nodata&14.0\p0.4&10.1\p0.4&1.11\p0.02& \nodata   \\
Field              & \nodata  &\nodata&\nodata&12.5\p1.0&10.6\p1.1&1.18\p0.04&22.8\p1.3\a\\
\enddata
\tablenotetext{a}{Derived from the Kennicutt (1998) sample of nearby galaxies.}
\label{tab:results}
\end{deluxetable}
\normalsize

\scriptsize
\begin {deluxetable}{ccccccc}
\def\a{\tablenotemark{a}}
\def\b{\tablenotemark{b}}
\tablewidth{425pt}
\tablecaption{Tully-Fisher Relation}
\tablehead{
\colhead{Cluster} & \colhead{$N_{\rm TF}$} & \colhead{$a_{\rm bias}-a_{\rm TF}$} & \colhead{$a-a_{\rm TF}$} & \colhead{$\sigma_a$} & \colhead{c$z_{\rm cmb}$} & \colhead{$v_{\rm pec,cmb}$}
\\
\colhead{} & \colhead{} & \colhead{(mag)} & \colhead{(mag)} & \colhead{(mag)} & \colhead{(\kmsb)} & \colhead{(\kmsb)}
}
\startdata
A2029 & 14 & $-$0.100 & $-$0.052\p0.157 & 0.53 (0.34) & 23400 & ~$+$550\p1700\\ 
A2218 & ~8 & $+$0.237 & $+$0.251\p0.138 & 0.34 (0.29) & 52497 & $-$6400\p3300\\
A2295 & 10 & $-$0.042 & $-$0.026\p0.137 & 0.38 (0.32) & 24555 & ~$+$290\p1500\\
A2670 & 11 & $+$0.003 & $+$0.012\p0.116 & 0.32 (0.17) & 22494 & ~$-$120\p1200\\
\enddata
\tablecomments{
Col. 3: Tully-Fisher offset with respect to the $z\sim0.1$ template zero point, uncorrected for cluster population incompleteness bias.
Col. 4: Tully-Fisher offset with respect to the $z\sim0.1$ template zero point, corrected for cluster population incompleteness bias.
Col. 5: Tully-Fisher scatter, with the intrinsic contribution in parentheses.
}
\label{tab:TF_results}
\end{deluxetable}
\normalsize

\centerline{\psfig{figure=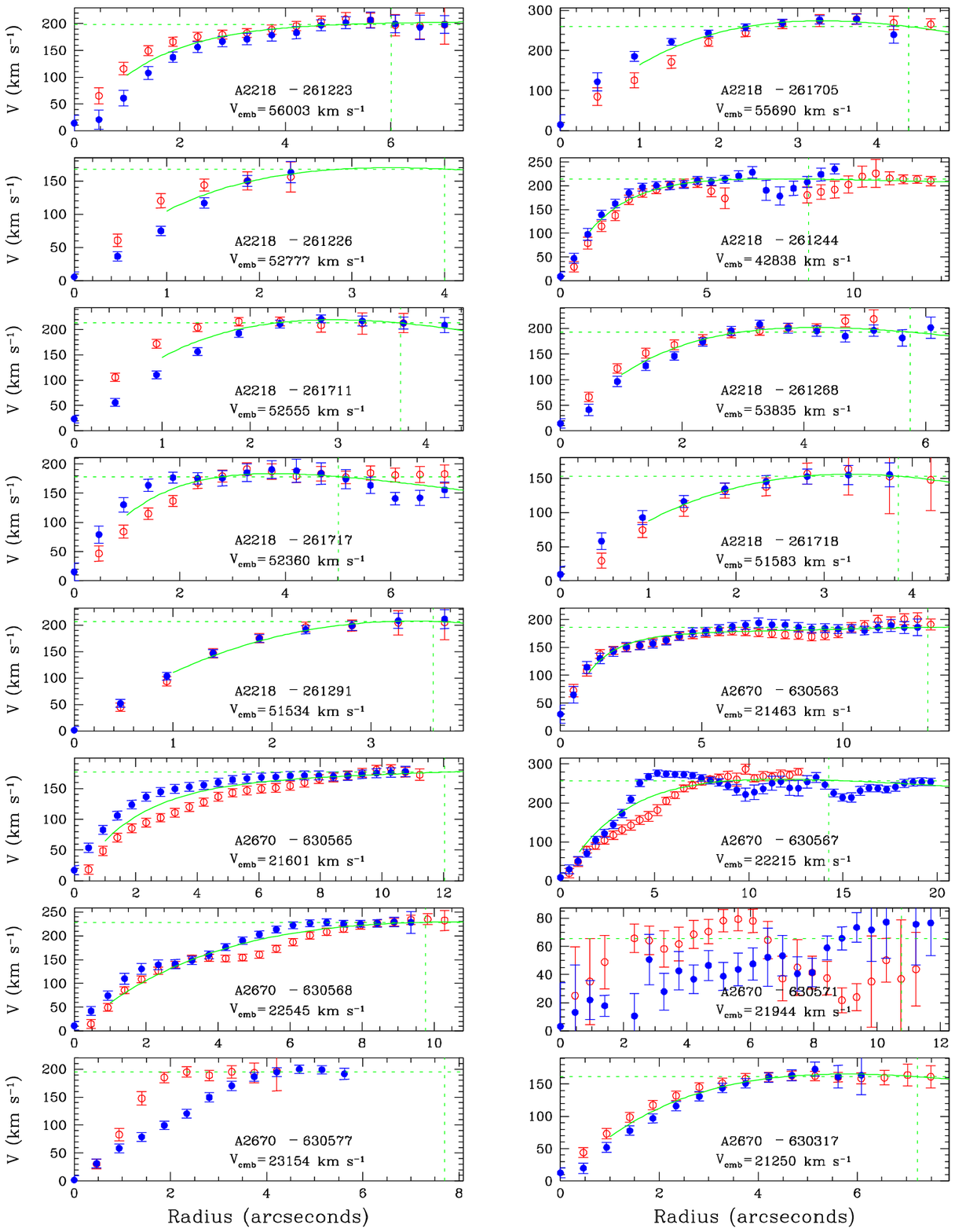,width=7.0in,bbllx=67pt,bblly=144pt,bburx=515pt,bbury=714pt}}
\newpage
\end{document}